\documentstyle[prd,epsfig,aps]{revtex}
\newcommand{\be}{\begin{eqnarray}}
\newcommand{\ee}{\end{eqnarray}}
\newcommand{\bi}{\bibitem}

\newcommand{\rar}{\rightarrow}

\newcommand{\mplq}{m_{Pl}^2}
\newcommand{\gmn}{g_{\mu\nu}}
\newcommand{\rmn}{R_{\mu\nu}}
\newcommand{\da}{D_\alpha}
\newcommand{\db}{D_\beta}
\newcommand{\dm}{D_{\mu}}
\newcommand{\dn}{D_{\nu}}
\newcommand{\cB}{{\cal B}}
\newcommand{\cC}{{\cal C}}

\begin{document}
\draft
\input epsf

\twocolumn[\hsize\textwidth\columnwidth\hsize\csname@twocolumnfalse\endcsname
\title{Determinant-Gravity: Cosmological implications }

\author{D. Comelli $^{(1)}$, A. Dolgov $^{(1,2)}$ }
\address{$^{(1)}${\it INFN - Sezione di Ferrara, 
via Paradiso 12, I-35131 Ferrara, Italy  }}
\address{$^{(2)}${ \it
ICTP,
Strada Costiera 11,
31014 Trieste, Italy,\\
INFN, sezione di Ferrara,
Via Paradiso, 12 - 44100 Ferrara,
Italy \\
ITEP, Bol. Cheremushkinskaya 25, Moscow 113259, Russia. }}
\date{April 2004}
\maketitle
\noindent

\vspace{0.5cm}

We analyze the action $\int d^4x\, \sqrt{\det||{\cal B}\,g_{\mu\nu}+
{\cal C} \,R_{\mu\nu}}||$ as a possible alternative or addition to the 
Einstein gravity. Choosing a particular form of ${\cal B}(R)= \sqrt {R}$ 
we can restore the Einstein gravity and, if ${\cal B}=m^2$, we obtain 
the cosmological constant term. Taking ${\cal B} = m^2 + {\cal B}_1 R$ 
and expanding the action in $ 1/m^2$, we obtain as a leading term the  
Einstein Lagrangian with a cosmological constant proportional to $m^4$ 
and a series of higher order operators. In general case of non-vanishing 
${\cal B}$ and ${\cal C}$ new cosmological solutions for the 
Robertson-Walker metric are obtained.


\vskip1pc

]

\vspace{2.3cm}

{\it 1. Introduction.}~~
There are numerous suggestions in the literature for modification of the
classical Einstein action of general relativity:
\be
S_E = \frac{\mplq}{16\pi}\,\int d^4x\, \sqrt g\, \left( R-2 \Lambda \right)
\label{se}
\ee
where $R$ is the curvature scalar, $g = - \det| g_{\mu\nu}|$ is the determinant
of the metric tensor, $\Lambda$ is the cosmological constant
and $m_{Pl}$ is the Planck mass. Majority of such  attempts
(at least in 4-dimensional space-time) are based on the same structure of the
action integral with an addition to the Einstein term of some
scalar functions of $R$ and/or of combinations of
the Ricci, $R_{\mu\nu}$, and/or Riemann, $R_{\mu\nu\alpha\beta}\,$, tensors
($ \int d^4x\, \sqrt g\, {\cal L}(R,R_{\mu\nu},R_{\mu\nu\alpha\beta} )$).
Usually, but not necessarily, one considers quadratic terms in the 
curvature 
proportional to $R^2$, $R_{\mu\nu}R^{\mu\nu}$, and 
$R_{\mu\nu\alpha\beta}R^{\mu\nu\alpha\beta}$. Such terms do not introduce a
scale dependent parameter or, in other words, they enter into the action
with dimensionless coefficients. 
Some of such and higher order or even non-local
terms may appear as a result of quantum 
corrections, see e.g. the book\cite{bd}.

Such a form of the Lagrangian density, i.e. a scalar function multiplied by
the determinant of the metric tensor, 
is dictated by the demand of the invariance of the action with 
respect to general coordinate transformations. However, this is not the
only way to ensure this invariance. In fact any scalar function multiplied
by a determinant of a second rank tensor would also be invariant with
respect to the choice of coordinates \cite{soko}.

Already in 1934 Born and Infeld \cite{born} proposed a new action for 
electromagnetism (reducing to the Maxwell action for small amplitudes)
 of  the form
\be
S_{BI}
=\int d^4 x \sqrt{  \det|| m^2\;g_{\mu\nu}+ {\cal F}_{\mu\nu}|| }
\label{bi}
\ee
with $ {\cal F}_{\mu\nu}$ the electromagnetic field strength.

A generalization of the above action to gravity was suggested by 
Deser and Gibbons in ref.\cite{deser}, where they
 studied the physical criteria that such a theory should satisfy.

Along the same  line 
we analyze 
maximally symmetric spaces where
only two independent tensors $g_{\mu\nu}$ and $R_{\mu\nu}$ 
are needed to specify geometrical structure of space-time.
Within such a hypothesis we can write the general covariant action as:
\be
{\cal S}_{det}= \int d^4 x\sqrt{  \det||  {\cal G}_{\mu\nu}|| }
\label{new}
\ee
where $ {\cal G}_{\mu\nu} =  {\cal B} \,g_{\mu\nu}+{\cal C}\,R_{\mu\nu}$
\footnote{ Such an ensemble of actions is more
 restricted than the one proposed by 
Deser and Gibbons \cite{deser} and does not satisfy the 
condition of absence of ghosts
in the linearized version \cite{deser},\cite{whol}.}.
Below we make a simple choice for the coefficients:
${\cal B}= {\cal B} (R) = m^2 + {\cal B}_1 R$ where $m$,
at this stage\footnote{The Planck mass scale  enters the 
Lagrangian only when we  define the  coupling between matter and gravity.
},  
is a new  mass parameter, while
${\cal B}_1$ and ${\cal C}$ are some dimensionless constant coefficients.
In general, ${\cal B}$ and ${\cal C}$ could be arbitrary scalar functions 
of geometrical tensors, $R$, $\rmn$, $R_{\mu\nu\alpha\beta}$, etc.

Somewhat similar modification with ${\cal L} = \sqrt{\det ||R_{\mu\nu}||}$
was suggested by  Eddington  \cite{eddi}, as we have found out from
ref.~\cite{magnano}, but the latter was considered as a pure affine theory
depending on connection $\Gamma^\alpha_{\mu\nu}$ and its derivatives only
( a generalization of the above results in the same framework
 is given by  Vollick \cite{volli}), 
while we prefer to
 remain in the frameworks of metric theory in the same spirit as 
ref.\cite{deser}.

It is non-trivial task to derive equations of motion from action 
(\ref{new}), 
but in the case of maximally symmetric space, considered here, this 
problem is greatly simplified. One can show that:
\be \nonumber
\frac{\delta{\cal S}_{det} }{\delta\, g^{\mu\nu}}&=&
-{\cal B}(R){\cal P}_{\mu\nu}+{\cal B}'\,{\cal P} \,R_{\mu\nu}
+\left( \gmn D^2 -
\dm\dn \right)
\\
&&\nonumber
 \left({\cal B}'{\cal P}\right)+
({\cal C}/2) \left(D^2 {\cal P}_{\mu\nu} +
\gmn \da\db {\cal P}^{\alpha\beta}-\right. \\
&&
\left.\da\dm {\cal P}^\alpha_\nu -
\da\dn {\cal P}^\alpha_\mu \right)=0
\label{eq-of-mot}
\ee
where ${\cal B}' = d{\cal B}/dR$,
$\dm$ is the covariant derivative
and $D^2 =\gmn D^\mu D^\nu$.

The second rank tensor ${\cal P}_{\mu\nu}$ is defined as
\be\label{pp}
{\cal P}_{\mu\nu} &=& \frac{\sqrt{\det|| {\cal G}_{\mu\nu}|| }}{2\, \sqrt g}\;
{\cal G}_{\mu\nu}^{(-1)}
\ee
and ${\cal P}=g^{\mu\nu}{\cal P}_{\mu\nu}$.

Remember that, according to our hypothesis, any geometric second rank tensor,
in particular, ${\cal P}_{\mu\nu}$
can be expressed through a linear combination of two basic tensors:
$ a_1\,g_{\mu\nu}+a_2\,R_{\mu\nu}$,
where $a_{1,2}$ are scalar functions
of various invariants ($R^n,\,\,Tr [ R \cdot ...\cdot R]$).

We checked explicitly that eq. (\ref{eq-of-mot}) satisfies, as it should,
the transversality condition, $D^\mu (l.h.s.) = 0$.

The physical consequences of the suggested above gravitational
action can be studied from different angles.

In the most conservative  approach 
we may consider the new action as a higher order correction
to the classical Einstein term and restrict
the new interactions with phenomenological considerations.
In this case we
have to put into the r.h.s. of the equation
of motion (\ref{eq-of-mot}), the usual general relativity term,
$8\pi T_{\mu\nu}/m_{Pl}^2 - G_{\mu\nu}$,
where $T_{\mu\nu}$ is the energy-momentum tensor of matter and
$G_{\mu\nu} = R_{\mu\nu} -(1/2)\gmn R$ is the Einstein tensor.
In the most radical  approach, instead, we can   work only with the 
new action as a fundamental dynamical source  of the gravitational field
 and may even try to introduce matter fields
under the signs of $\det$ and the square root (work in progress).

In such a perspective it is important, first of all, 
to find out the proper limiting connection to the classical approach,
eq.(\ref{se}), trying to reconstruct the Einstein General Relativity 
limit using a special form of
functions $\cal{B}$ and $\cal{C}$ (see below).
 
The first problem can be solved if we take 
$\cB(R) = m^2 + \cB_1 R$, as we 
mentioned above, and expand the  action (\ref{new}) 
in powers of $1/m^2$
\be
S_{det} &\approx& \int d^4 x \sqrt g \,\left[ m^4+ (2\,{\cal B}_1 + 
\right.
{\cal C}/2) m^2 \,R+\\
&&\nonumber
({\cal B}_1^2+{\cal B}_1{ \cal C}/2) R^2 +
\left.
{\cal C}^2/4\,[R]^2+...\right]
\ee
where $[R]^2=R_{\mu\nu}R^{\mu\nu}$. The
first and second  terms in the expression above correspond to the Einstein 
action with a cosmological constant
(see eq. (\ref{se})) while the higher order terms are the corrections to
be treated perturbatively.
It is clear that, if we want to get rid of the cosmological constant,
in the radical approach, we cannot send 
naively $m\rightarrow 0$ because we lose the classical limit,
hence we have to take into account 
canceling matter effect (work in progress)
\footnote{A naive cancellation mechanism consist in the addition
 to eq. (\ref{new}) of the term 
$-\int d^4 x\; m^4\,\sqrt{\det||g_{\mu\nu}||}$ }.

\vspace{0.3cm}

We will study eq.~(\ref{eq-of-mot}) in homogeneous and isotropic space-time
with the FRW metric: 
\be
ds^2 = dt^2 - a^2(t) dr^2.
\label{ds2}
\ee
As is well known it is sufficient in this case
to consider only the time-time component of equations of motion.

The explicit general solution of the above equation is quite cumbersome
(and above the length of our paper) so we will try to find some 
results in physically interesting simple limits.

\begin{itemize}

\item 
For example it is easy to see that
a De Sitter phase ($a(t)=e^{h\,t}$) is a solution of eq.(\ref{eq-of-mot})
of  the general action (\ref{new}) if
\be
m^2\left(m^2-3(4\,\cB_1+\cC)\,h^2\right)=0
\ee
and in order to have  real $h$ solutions, 
the parameters $\cB_1$ and $\cC $ should lie in certain bounds. 
In particular there are  solutions with fixed  $h^2=m^2/(
\sqrt{3}(\cC+4 \cB_1))$ if $\cC+4\,\cB_1>0$ and $m\neq 0$ but we 
found that they are instable under small
perturbations. On the contrary, 
if $m=0$, there are solutions with generic   $h\neq 0$
(and we checked their stability for any $\cB_1,\cC$).

\item The study of possible power law solutions of 
eqs.(\ref{eq-of-mot}), $a(t)=t^n$, 
is easier in the scale invariant case where $m=0$ and $\cB=\cB_1\,R$.
Note  that  the new action containing
 $\det|| \cB_1 \,R\,\gmn+\cC\, \rmn ||$
 can be meaningful only for a definite sign of the determinant.
 In contrast to $\det|| \gmn||$ the latter is
not guaranteed from the general principles but the sign may be 
fixed on the solutions of equations of motion.
In fig.\ref{plusminus} we show the signature of the determinant  versus 
 the ratio
$\cC/\cB_1$ and $n=2/(3+3\,w)$. We see that 
a well defined signature in the interval $-1<w<1$ is realized
 only for 
$\cC/\cB_1=-4,\,0 $: the second solution is, in a sense, ``trivial''
corresponding to the well known quadratic 
$R^2-$Lagrangian, $\cB_1^2\,\int d^4x \sqrt{g}\, \,R^2 $, while
the first solutions generates the traceless tensor 
$ {\cal G}_{\mu\nu}=\cB_1( R\,g_{\mu\nu}-4\,R_{\mu\nu})$ 
whose equations of motion are studied below.
\begin{figure}[htb]
\centering
\epsfxsize=2.8 in
\epsfysize=2.6 in
\begin{center}
\leavevmode
 \epsfbox{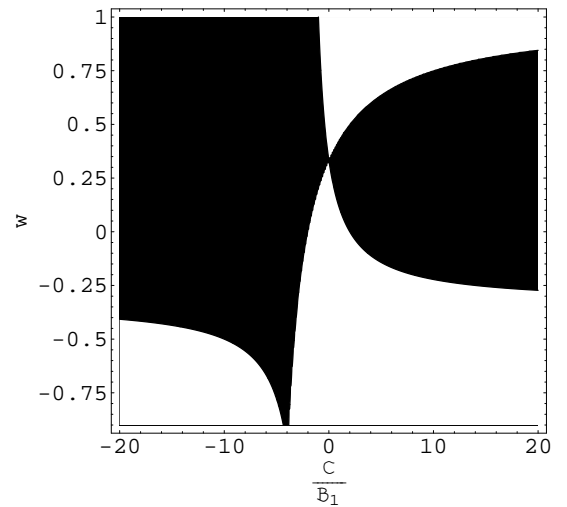}
\end{center}
\caption{
Signature of
$\det|| \cB_1\, R\,g_{\mu\nu}+\cC\, \rmn ||$
for $a(t)=t^n$ and  $n=2/(3+3w)$.
The negative values are shown as black areas, while the positive ones
are whites.}
\label{plusminus}
\end{figure}

For the power law solutions, $a(t)=t^n$, we find the constraint
\be\label{ww}
w=-\frac{1}{3}+\frac{8(3+x)}{3(2+x)(6+x)}
\ee
where $n=2/(3+3\,w)$, $x=\cC/\cB_1$.

  This curve is showed in fig.\ref{wab}.
In order to control stability of the solutions
we have found that the first curve to the left corresponds 
to unstable solutions.
For the central one   there is a stability region for 
$0\leq w<-1$ corresponding to $-2\sqrt{3}\leq x<2(-3+\sqrt{3})$, and 
for the third curve to the right,
stability is ensured if $-1/3<w\leq 0$ and $x\geq 2\sqrt{3}$. Thus,
there is no stable solution for $w>0$ and the matter dominated 
regime ($w=0$) is the stability border line.

For example the  solution $a(t)=t^{1/2}$ corresponding to $w=1/3$ 
(radiation dominated universe)
are obtained for $x=-4$ and for $x=0$
and both results are unstable.
\begin{figure}[htb]
\epsfxsize=2.8 in
\epsfysize=2.4 in
\begin{center}
\leavevmode
 \epsfbox{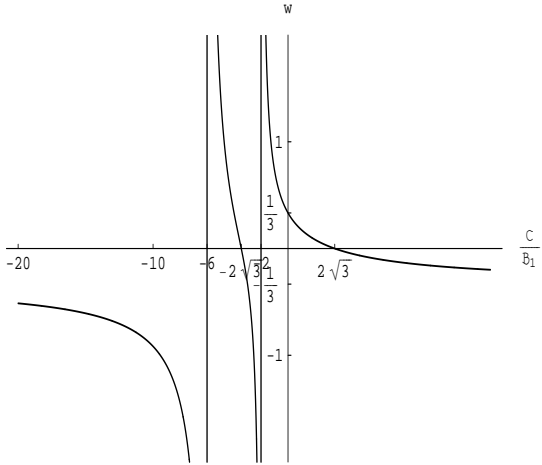}
\end{center}
\caption{
Plot of  eq.(\ref{ww}) which gives the values of $w$
as function of $\cC/\cB_1$ indicating when 
the power law  $a(t)=t^{2/(3+3w)}$ ansatzs
is a solution of the scale invariant Lagrangian with $m=0$.
}
\label{wab}
\end{figure}
\item 

The minimal scale invariant case with  $m=\cB_1=0$
\be
S_{\det}\rightarrow \cC^2\,\int d^4 x \sqrt{det||R_{\mu\nu}||} 
 \label{detr} 
\ee
gives a tractable equation of motion:
\be
 \cC^2\,H^4\! \sqrt{\frac{3 +\chi}{3(1+\chi)}}\!\!
\left[\frac{\dot\chi}{H}\,\frac{2\chi^2+6\chi+3}{(3\! +\!\chi)(1\!+\!\chi)}
+ \!3\chi (1\!+\!\chi)\! \right]\! =\! 0
\label{eq-cosm}
\ee
where $H=\dot a/a$ is the Hubble parameter and $\chi = \dot H/H^2$.
\begin{figure}[htb]
\epsfxsize=3.1 in
\epsfysize=2. in
\begin{center}
\leavevmode
 \epsfbox{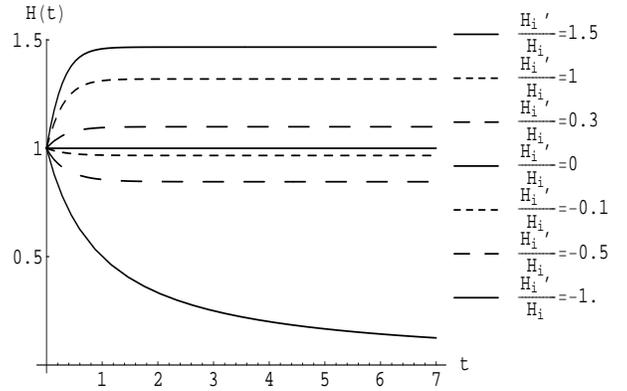}
\end{center}
\caption{Solutions of eq.(\ref{eq-cosm}) in the case $H_i\geq 0$,
where  $H(t)$ is evolving for different initial
conditions $H_{i}'/H_{i}=1.5,1,0.3,0,-0.1,-0.5,-1 \;$ and $H_i=1$ .
}
\label{minimal1}
\end{figure}
At expansion regime, when $H>0$ the solutions tend to the 
De Sitter one with $H\rar const$ both for negative and positive initial
values of $\dot H$ 
(see fig.\ref{minimal1}), while at contraction, 
when $H<0$, the solutions may tend
to singularity, $H\rar -\infty$ if $\dot H_{i} <0$ or  
to $H=0$ if the initial value of $\dot H_{i}$ is positive
(see fig.\ref{minimal2}).
However,
after reaching zero value. $H$ does not change sign but becomes negative
again and also reaches singularity. Thus, in this approximation 
singularity is not avoided, though the contraction may stop for a while.
\begin{figure}[htb]
\begin{center}
\leavevmode
\epsfxsize=3.1 in
\epsfysize=2.0 in
 \epsfbox{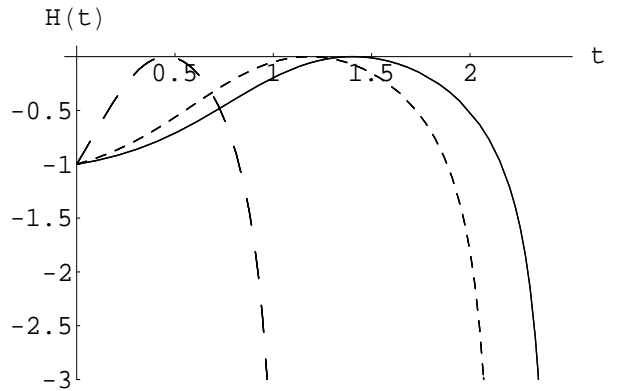}
\end{center}
\caption{Solutions of eq.(\ref{eq-cosm}) in the case $H_i\leq 0$,
where  $H(t)$ is evolving for different initial
conditions $H_{i}'/H_{i}=\;0.3,\,0.5\,,3$ and $H_i=-1$ .
}
\label{minimal2}
\end{figure}

\item
The scale invariant case with $m=0$ and $\cC=-4\,\cB_1$ also
deserves some comments. The action in this case is:
\be
S_{\det}\rightarrow  \cC^2\,\int d^4 x \sqrt{det||
R\,g_{\mu\nu}
-4\,R_{\mu\nu}||}
\label{Tr0}  
\ee
The determinant of ${\cal G}_{\mu\nu}$ is sign definite and the ratio
$\det {\cal G}_{\mu\nu} /\det g_{\mu\nu} = (2\dot H)^4$. 
The equation of motion is very simple:
\be
2 \ddot H H + 6 H^2 \dot H - \dot H^2  =0
\label{0-trace-eq}
\ee
and can be integrated analytically in quadratures but
numerical solution is faster and simpler. At expansion regime, when
initially $H_{i}>0$, the solution tends to a De Sitter 
one with $H\rar const$.
On contraction phase, when $H_{i}<0$, we find that either $H$ reaches 
negative and infinitely large value in finite time, 
if $\dot H_{i} <0$, or
$H$ tends to zero, again in finite time if $\dot H_{i}>0$.

Surprisingly equation (\ref{0-trace-eq}) is the same that can be derived 
from the simple quadratic Lagrangian $\int d^4x\,\sqrt{g}\; R^2$ \cite{R2} or,
 in our approach, from action (\ref{new}) with $m=0$ and ${\cal C} = 0$.

It is interesting to study eq.(\ref{0-trace-eq})
in presence of the Einstein term and matter,
that means we have to add the term 
$ \kappa (T_{00}- G_{00} )$ to the right hand side,
where $\kappa$ is a constant, $T_{00}=\rho$ is the energy density
of matter, and $G_{00} = 3 H^2$. Here we took the system of units with 
$m_{Pl}^2/8\pi =1$.

In this case the behavior of the solution 
at expansion regime becomes different. If $\kappa >0$ the solution 
$H(t)$ has an oscillating behavior. For vacuum energy, when 
$\rho = const$, the oscillations
proceed around a constant average value (see fig.\ref{matlamb}), 
while for normal matter with
$\rho \sim a^{-n}$, where $n=3$ for non-relativistic matter and $n=4$ for
relativistic matter, the oscillations proceed around gradually decreasing 
average values (see fig.\ref{matlamb}). The acceleration parameter, 
$(\ddot a/a) = \dot H + H^2$
also oscillates and may be both negative and positive, depending upon
time moment. The line $H=0$ is a separation line
of equation (\ref{0-trace-eq}) 
and solutions can smoothly touch it but never cross.
\begin{figure}[htb]
\epsfxsize=2.8 in
\epsfysize=2. in
\begin{center}
\leavevmode
 \epsfbox{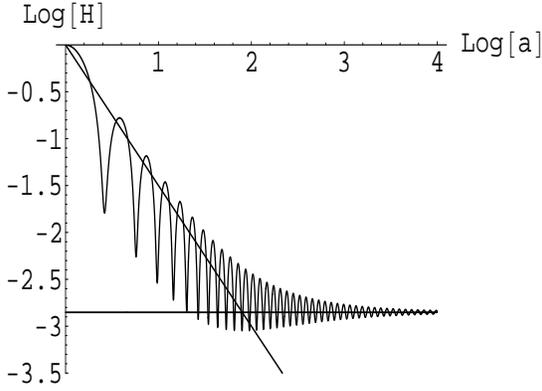}
\end{center}
\caption{Solutions of eq.(\ref{0-trace-eq}) in 
presence of matter and cosmological constant ($\rho=\rho_{matter}+
\rho_{\Lambda}=e^{-3 z}+0.01$)
($z=\log[a(t)]$)
in  the case $H_i= 1,\,H'_i=0, \,\kappa=10$.
We plot the value of $\log[H]$ and the descending
 straight line represents  $\log[\rho_{matter}]=-3\,z$ 
while the horizontal one is the attraction point
$3H^2= \rho_{\Lambda}$
}
\label{matlamb}
\end{figure}

\end{itemize}

As we have already mentioned, the determinant of ${\cal G}_{\mu\nu}$ is
sign definite only for ${\cal C}/{\cal B}_1 =0,-4$. In other cases it may
change sign in the course of evolution and the solution would encounter
quite strange singularity. To avoid that we may try the action with
the absolute value, ${|\det\cal G}_{\mu\nu}|$. It preserves invariance
with respect to coordinate transformation, since the latter does not
change the sign of determinant of a second rank tensor. Equations of 
motion in this case acquire an additional factor containing
the sign function, $\epsilon [z] = z/|z|$. 
In particular the only change in eq. (\ref{eq-of-mot}) is the change 
of ${\cal P}_{\mu\nu}$ by the absolute value under the sign of the
square root.

As a result the
singularity, at the point where the determinant vanishes, disappears
and the equations of motions can be solved for all values of time.
We have found some solutions of time-time component of 
the equations of
motion for a particular case of ${\cal G}_{\mu\nu}$ coinciding with
the Einstein tensor ${\cal G}_{\mu\nu}= {\cal C}(\rmn - \gmn R/2) $.

In this particular case, the equation for time-time component has a very 
simple form:
\be
\ddot H = - 3 \dot H\,H + \kappa  \sqrt{|2\dot H + 3 H^2 |}\,
\epsilon [2\dot H + 3 H^2 ]\,\frac{ 3H^2 -\rho}{H|H|}
\label{ein-eq}
\ee
where $\kappa$ is a constant which may be both positive and negative, 
$\epsilon [z]$ is the sign function as has been already mentioned above.
This sign function appears as a result of taking the absolute value of the
determinant of ${\cal G}_{\mu\nu}$, while $|H|$ 
appears because this determinant
is proportional to $\sqrt{H^2}=|H|$.

\begin{figure}[htb]
\epsfxsize=3.2 in
\epsfysize=2.4 in
\begin{center}
\leavevmode
 \epsfbox{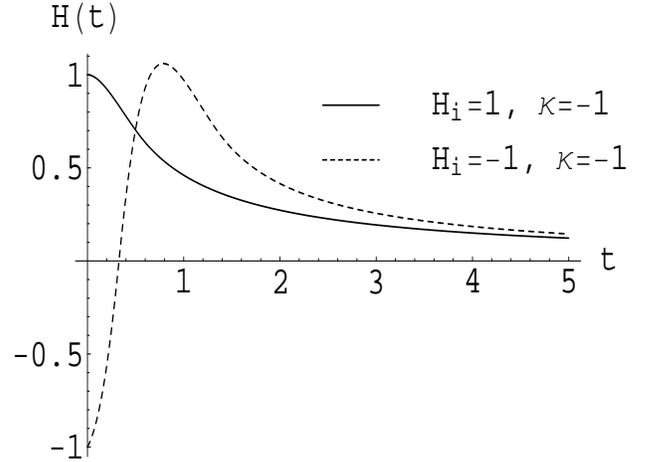}
\end{center}
\caption{Solutions of eq.(\ref{ein-eq}) in absence of matter ($\rho=0$)
 and with the initial conditions $H_i= \pm 1,\,H'_i=0, \kappa=-1$.
}
\label{puregrav}
\end{figure}

We have solved this equation for different special cases. 
At expansion regime,
when $H>0$, the solution without matter tends to $H= 2/3t$, i.e. to
the matter dominated (MD) regime but
without matter (see fig.\ref{puregrav})). 

\begin{figure}[htb]
\epsfxsize=3.2 in
\epsfysize=2.2 in
\begin{center}
\leavevmode
 \epsfbox{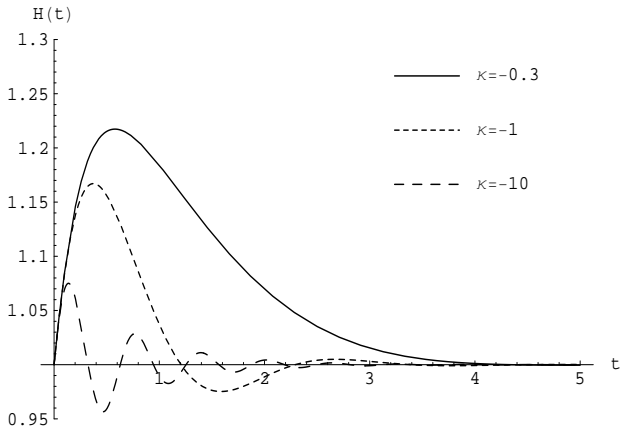}
\end{center}
\caption{Solutions of eq.(\ref{ein-eq}) in presence of the 
cosmological constant, $\rho_{vac}=3$,
and with the initial conditions $H_i=  1,\,H'_i=1$ for different 
values of $\kappa=-0.3,\,-0.1,\,-10$.
}
\label{matt+grav}
\end{figure}

The same behavior is found in presence
of relativistic and non-relativistic matter. If vacuum energy is non-zero
and initially sub-dominant then the solution starts from MD regime and then
turns into De Sitter one with $H \rar const$. An interesting feature
of this solution is that before reaching an asymptotic constant value the 
solution oscillates around it with the frequency 
$\omega^2 \approx \kappa m_{Pl}^2 /8\pi$ (see fig.\ref{matt+grav}). 
Such fast 
oscillations would give rise
to gravitational particle production with energies near the Planck energy.
This process might be effective at the early inflationary
 stage and probably even
now when $H^2 m_{Pl}^2$ dropped down to the present day 
value of the vacuum energy.
It is tempting to identify these gravitationally produced particles with the
sources of ultra-high energy cosmic rays.

For the case of initially negative $H$ (contraction regime) the solution in
the absence of matter and vacuum energies demonstrates very interesting 
feature
that the Hubble parameter grows up to zero, becomes positive and after reaching
some maximum value turns to MD regime. Thus in the ``empty'' universe 
cosmological solutions always turn into expanding ones with $H=2/3t$ even if 
started from contraction. In this simple case cosmology is non-singular 
and contraction turns into expansion (see fig.\ref{puregrav})). 

\begin{figure}[htb]
\epsfxsize=3.2 in
\epsfysize=2.2 in
\begin{center}
\leavevmode
 \epsfbox{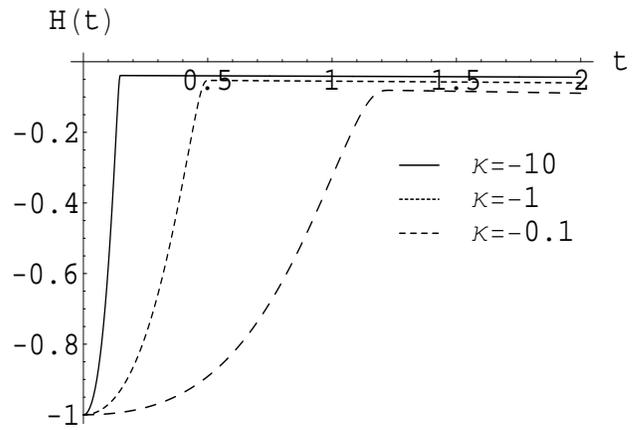}
\end{center}
\caption{Solutions of eq.(\ref{ein-eq}) in presence of matter density
 ($\rho=0.1/a(t)^3,\;\;a(0)=1$)
with the initial conditions $H_i= -1,\,H'_i=0$ and different values of
$\kappa=-0.1,\,-1,\,-10$.
}
\label{absmatt}
\end{figure}

More interesting and more realistic situation is non-vanishing 
energy density of
matter and/or vacuum. In this case the line $H=0$ cannot be crossed, 
at least as
we were able to check on some simple examples
(see fig. \ref{absmatt}). We have not found any realistic
case of non-singular contraction but cannot exclude that such may exist.

This approach can be also implemented in  multidimensional case \cite{nieto}.
It is worth to note that a change in number of dimensions 
does not request a change in the dimensionality of the coefficients in the
action.

As any other higher derivative modifications of gravity, the theory 
introduced here possibly encounters all the problems which
such theories are known to have in quantum case \cite{renormaliz},
 \cite{deser}.
 Still
more detailed investigation seems necessary.

\acknowledgements
We thank S.~Deser and G.~W.~Gibbons for informing us 
about their paper \cite{deser}.

\end{document}